\documentclass[aps,prl,twocolumn,groupedaddress,showpacs,showkeys]{revtex4-1}%
\usepackage{amsfonts}
\usepackage{mathtools}
\usepackage{amsmath}
\usepackage{amssymb}
\usepackage{float}
\usepackage{subdepth}
\usepackage{color}
\usepackage{graphicx}
\usepackage{natbib}
\usepackage{hyperref}
\usepackage{url}
\usepackage{soul}%
\providecommand{\U}[1]{\protect\rule{.1in}{.1in}}
\hypersetup{colorlinks, linkcolor={blue}, citecolor={blue}, urlcolor={blue}}
\begin{document}

\title{The sliding phase transition in ferroelectric van der Waals bilayers}
\author{Ping Tang$^{1}$}
\author{Gerrit E. W. Bauer$^{1,2,3,4,5}$}
\affiliation{$^1$WPI-AIMR, Tohoku
University, 2-1-1 Katahira, Sendai 980-8577, Japan}
\affiliation{$^2$Institute for Materials Research, Tohoku University,
2-1-1 Katahira, Sendai 980-8577, Japan} \affiliation{$^3$Center for
Spintronics Research Network, Tohoku University, Sendai 980-8577, Japan}
\affiliation{$^4$Zernike Institute for Advanced Materials, University of
Groningen, 9747 AG Groningen, Netherlands}
\affiliation{$^5$Kavli Institute for Theoretical Sciences, University of the Chinese Academy of Sciences,
Beijing 10090, China}

\begin{abstract}
We address the sliding thermodynamics of van der Waals-bonded bilayers by the continuum elasticity theory. We attribute the robustness of the ferroelectricity recently observed in h-BN and WTe$_{2}$ bilayers to large monolayer in-plane stiffness. We compute the electric susceptibility and specific heat in the mean-field self-consistent phonon approximation. We
compare critical temperatures and electric switching fields with the observations.
\end{abstract}
\maketitle

The discovery of ferroelectricity in van der Waals stacked bilayers of
two-dimensional (2D) WTe$_{2}$ and hexagonal boron nitride (h-BN) with
out-of-plane polarization substantially expands the family of ferroelectric
materials \cite{Li2017,Fei2018a,Yang2018,Liu2019, Sharma2019, Wu2021,  Yasuda2021,Stern2021,Wang2022}. The dipolar order arises from the precise
stacking of two polar van der Waals-bonded monolayers that change sign by a
small shear motion. The potential barriers for switching between the up and
down polarization states are very low ($\lesssim$ meV per unit cell)
\cite{Li2017, Yang2018, Wu2021}. Surprisingly, the \textquotedblleft sliding
ferroelectricity\textquotedblright\ remains stable even above room temperature
\cite{Fei2018a, Yasuda2021, Stern2021, Wang2022}, in contrast to the ferromagnetism in van der Waals mono or bilayers
 \cite{Lee2016,Huang2017,Gong2017,Fei2018b,Bonilla2018,OHara2018,
Deng2018,Huang2018b}.

From a theoretical perspective, long-range order weakens with reduced
dimensionality ($d$) \cite{Li2021}. According to the Mermin-Wagner-Hohenberg
theorem \cite{Mermin1966,Hohenberg1967} at any finite temperatures an
isotropic short-range force cannot order spin system with $d\leq2$ due to the
infrared divergence caused by gapless Goldstone modes. An anisotropy or a
switching barrier is thus essential for phase transitions in $d\leq2$. 2D
magnets are stable at room temperature only when the magnetic anisotropy
amounts to tens of meV per magnetic moment. The mechanism underlying the high
thermal stability of sliding ferroelectrics in spite of the low switching
barriers appears to be unexplained.

In this Letter we present a thermodynamic model of 2D sliding ferroelectrics
that explains this conundrum. We associate the sliding ferroelectric phase
transition with the shear motion of the entire layer with macroscopic mass that is driven by thermally fluctuating forces. The
model parameters include the mass density, intralayer stiffness, and interlayer bonding. The phase transition is triggered by a soft \textquotedblleft sliding phonon\textquotedblright\ of the bilayers and the high Curie temperature follows from the interplay between the ultralow
switching barrier and intralayer rigidity. This mechanism is not
unique for ferroelectrics, but also holds for structural sliding instabilities
in non-ferroelectric bilayers, in which the phase transition can be observed
in the specific heat. However, the ferroelectricity serves as a unique monitor
of a bistability that can be controlled by temperature-dependent critical
switching fields.

\begin{figure}[ptb]
\centering
\par
\includegraphics[width=6.2 cm]{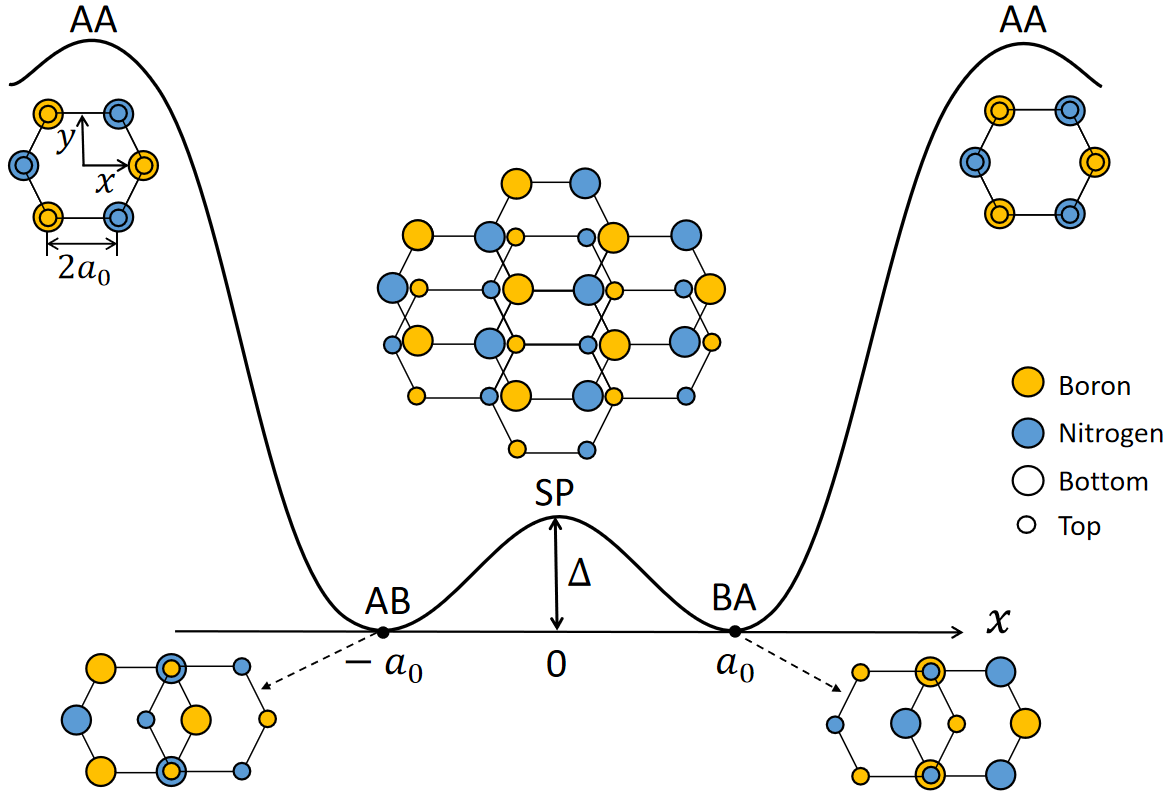}\newline\caption{The interlayer
binding energy landscape in sliding ferroelectrics illustrated for hexagonal
BN bilayers. The AB and BA stacking configurations correspond to two opposite
spontaneous polarization states that are separated by saddle-shaped potential
with minimum energy barrier $\Delta$ (per unit area). Boron and nitrogen atoms in the top
(bottom) layer are represented by large (small) orange and blue circles,
respectively. }%
\label{Fig-1}%
\end{figure}

We consider a bilayer of two atomic monolayers that may slide relative to each
other along a particular direction, e.g., the armchair (long lattice vector)
direction in the parallel stacked h-BN (WTe$_{2}$) bilayer. The energy minima correspond to states with opposite polarity that are separated by a saddle-point (SP) potential
barrier $(\Delta)$ defined by an intermediate non-polar configuration, as sketched in figure~\ref{Fig-1}. In the presence of a perpendicular electric
field $E$, the Hamiltonian of a bilayer under a relative sliding displacement
$\hat{u}_{s}$ along the $x$ direction reads \cite{Suzuura2002, Andres2012}
\begin{align}
\hat{\mathcal{H}}= &  \int\left [ \frac{\hat{\pi}_{s}^2}{2\rho_{s}}
+\frac{\lambda+2\mu}{2}\left(  \frac{\partial\hat{u}_{s}}{\partial x}\right)
^{2}+\frac{\mu}{2}\left(  \frac{\partial\hat{u}_{s}}{\partial y}\right)
^{2} \right\}  d^{2}\mathbf{r}\nonumber\\
&  +\int\left[  V_{B}\left(  \hat{u}_{s}\right) -EP\left(  \hat{u}%
_{s}\right)  \right]  d^{2}\mathbf{r},\label{Hal}%
\end{align}
where $\rho_{s}=\rho/2$ is half of the mass density $\rho$ of a single layer, $\hat{\pi}_{s}=\rho_{s}\dot{\hat{u}}_{s}$ the conjugate momentum to $\hat{u}_{s}$, $\lambda$ and $\mu$ the 2D Lam\'{e} coefficients, and $V_{B}$ the interlayer
binding energy density. $P(\hat{u}_{s})$ is the electric polarization density that depends on $\hat{u}_{s}$ and should be evaluated self-consistently below. Here we consider only one-component sliding motion, disregarding the interlayer displacements that do not directly affect the polar states such as out-of-plane flexural modes. We also neglect weak modulations of the electrostatic energy beyond the Stark interaction $-EP$.

\begin{table*}[ptb]
\caption{The parameter $\gamma$ and Curie temperature $T_{c}$ calculated for
several sliding bilayer ferroelectrics with model parameters extracted from
first-principles calculations \cite{Sachs2011, Torun2016, Li2017, Liu2019}.}%
\label{Table-1}%
\begin{ruledtabular}
\begin{tabular}{cccccccccccc}
&$\Delta$  &$\lambda $ &$\mu$ & $A_{0}$&$\rho$ &$a_{0}$& $P_{0}$ & $\gamma$& $T_{c}$\\
\hline
WTe$_2$ & $1.37\times 10^{-2}$ & $1.89$ & $2.69$ &21.8
& 68.18  & 0.246 & 0.38 & 1.69& 660\\
h-BN& $1.67$ & 3.37 & 7.67 &5.38
& 7.81 & 0.72 & 2.08 & 0.71&  $1.58\times 10^4$\\
Units& meV/\AA$^2$ & eV/\AA$^2$ & eV/\AA$^2$ & \AA$^2$
&$10^{-7}\,$kg/m$^2$ & \AA & pC/m & $10^{-2}$ & K\\
\end{tabular}
\end{ruledtabular}\end{table*}

$P(\hat{u}_{s})$ is an odd function of $\hat
{u}_{s}$ with respect to the non-polar SP, to leading order therefore
$P(\hat{u}_{s})=Z\hat{u}+\mathcal{O}(\hat{u}_{s}^{3})$, where $Z$ is a
constant that measures the interlayer polarization by the ionic charges. When
$Z=0$, the electric polarization and field effect vanish; our model then
describes a sliding structural phase transition between degenerate ground
states \cite{Palau2018}.

$V_{B}(\hat{u}_{s})$ is in general periodic for a large sliding distance. However, since the polar states are usually separated by a very low barrier and a small sliding displacement, we may adopt an approximate inverted camel-back potential 
 \cite{Yang2018}
\begin{equation}
V_{B}(\hat{u}_{s})=\frac{\Delta}{a_{0}^{4}}\left(  \hat{u}_{s}^{2}-a_{0}%
^{2}\right)  ^{2} \label{DB}%
\end{equation}
where $\Delta$ represents the barrier height per unit area and $2a_{0}$ is the
distance between the two minima. When $E=0,$ Eq.~(\ref{DB}) hosts two
degenerate minima at $\pm a_{0}$ with polarization $P_{0}=\pm Za_{0}$.

In general
\begin{equation}
\hat{u}_{s}(\mathbf{r},t)=\langle\hat{u}_{s}\rangle+\hat{\xi}_{s}%
(\mathbf{r},t) \label{flu}%
\end{equation}
where $\langle\cdots\rangle$ denotes the thermal average, $\hat{\xi}%
_{s}(\mathbf{r},t)$ are the spatio-temporal fluctuations with $\langle\hat{\xi}_{s}(\mathbf{r},t)\rangle=0$, and $\left\langle \hat{\xi
}_{s}^{n}(\mathbf{r},t)\right\rangle =\left\langle \hat{\xi}_{s}%
^{n}\right\rangle $ is independent of time and space. At equilibrium the force on each layer vanishes, i.e.,
\begin{align}
\langle\dot{\hat{\pi}}_{s}(\mathbf{r},t)\rangle=-\frac{i}{\hbar}\left\langle
[\hat{\pi}_{s}(\mathbf{r},t),\hat{\mathcal{H}}]\right\rangle=0. \label{EC}%
\end{align}
With Bosonic commutation relations $[\hat{\pi}_{s}(\mathbf{r},t),\hat{u}_{s}(\mathbf{r}^{\prime},t)]=-i\hbar \delta(\mathbf{r}%
-\mathbf{r}^{\prime})$ and $[\hat{\pi}_{s}(\mathbf{r},t),\hat{\pi
}_{s}(\mathbf{r}^{\prime},t)]=0$, Eq.~(\ref{EC}) leads to
\begin{equation}
\langle\hat{u}_{s}\rangle^{3}+3\langle\hat
{u}_{s}\rangle\langle\hat{\xi}_{s}^{2}\rangle+\langle\hat{\xi}_{s}^{3}%
\rangle-a_{0}^{2}\langle\hat{u}_{s}\rangle =\frac{Za_{0}^{4} E}{4\Delta } \label{ECo}%
\end{equation}
since the spatial gradient terms in Eq.~(\ref{Hal}) vanish on average. The dynamic equation for the fluctuations can be found from Heisenberg's equation of motion, $\dot{\hat{\pi}}_{s}=(-i/\hbar)[\hat{\pi
}_{s},\hat{\mathcal{H}}]$, as
\begin{align}
\rho_{s}\ddot{\hat{\xi}}_{s}  &  =\frac{\lambda+2\mu}{2}\frac
{\partial^{2}\hat{\xi}_{s}}{\partial x^{2}}+\frac{\mu}{2}\frac{\partial
^{2}\hat{\xi}_{s}}{\partial y}\nonumber\\
&  -\frac{4\Delta}{a_{0}^{4}}\left[  \left(  \langle\hat{u}_{s}\rangle
+\hat{\xi}_{s}\right)  ^{3}-a_{0}^{2}\left(  \langle\hat{u}_{s}\rangle
+\hat{\xi}_{s}\right)  \right]  +ZE. \label{SPH}%
\end{align}

We solve Eqs.~(\ref{ECo}) and (\ref{SPH}) in the self-consistent phonon scheme \cite{Lines1974,
Glass1976} using the mean-field approximations $\hat{\xi}_{s}^{2}%
\approx\langle\hat{\xi}_{s}^{2}\rangle$ and $\hat{\xi}_{s}^{3}\approx
3\langle\hat{\xi}_{s}^{2}\rangle\hat{\xi}_{s}$. Eq.~(\ref{ECo}) then reduces
to
\begin{equation}
\langle\hat{u}_{s}\rangle\left(  \langle\hat{u}%
_{s}\rangle^{2}+3\langle\hat{\xi}_{s}^{2}\rangle-a_{0}^{2}\right)  =\frac{Za_{0}^4 E}{4\Delta}.
\label{uT}%
\end{equation}
When $E=0,$ two roots are ferroelectric $\langle\hat{u}_{s}\rangle=\pm
(a_{0}^{2}-3\langle\hat{\xi}_{s}^{2}\rangle)^{1/2}$ and one is paraelectric
$\langle\hat{u}_{s}\rangle\equiv0$. With Eq.~(\ref{uT}), we can rewrite
Eq.~(\ref{SPH}) in the form of a harmonic oscillator in momentum space with
$\xi_{s}(\mathbf{q},t)=\int d^{2}\mathbf{r}\xi_{s}(\mathbf{r}%
,t)e^{-i\mathbf{q}\cdot\mathbf{r}}$
\begin{equation}
\ddot{\hat{\xi}}_{s}(\mathbf{q},t)=-\Omega_{\mathbf{q}}^{2}\hat{\xi}%
_{s}(\mathbf{q},t),
\end{equation}
with frequency dispersion that acquires a gap $\sim\sqrt{\Delta}:$ \textbf{ }
\begin{align}
\Omega_{\mathbf{q}}  &  =\frac{1}{\sqrt{\rho}}\left[  \frac{8\Delta}{a_{0}%
^{4}}\left(  3\langle\hat{u}_{s}\rangle^{2}+3\langle\hat{\xi}_{s}^{2}%
\rangle-a_{0}^{2}\right)  +(\lambda+2\mu)q_{x}^{2}\right. \nonumber\\
&  \left.  +\mu q_{y}^{2}\right]  ^{1/2}. \label{disp}%
\end{align}

Quantum mechanics enters the problem at low temperatures $T$ and high
frequencies when $\hbar\Omega_{\mathbf{q}}\gtrapprox k_{B}T,$ where $\hbar$
$\left(  k_{B}\right)  $ is Planck's (Boltzmann's) constant. The mean-square
of the fluctuations from the equilibrium position of an ensemble of harmonic
oscillators reads
\begin{align}
\langle\xi_{s}^{2}\rangle &  =\int\frac{\hbar}{\rho\Omega}%
\coth\left(  \frac{\hbar\Omega}{2k_{B}T}\right) D(\Omega) d\Omega. \label{OSC}%
\end{align}
where $D(\Omega)=1/(2\pi)^2\int d^{2}\mathbf{q}\delta (\Omega-\Omega_{\mathbf{q}})$ is the density of state of the sliding phonons. We regulate the divergence of the integral over $\Omega$ by a Debye frequency $\Omega_{D}$ cut-off chosen such that the degrees of freedom of the sliding motion per unit cell is conserved, i.e., $\int_{\Omega\leq\Omega_{D}}D(\Omega)d\Omega=1/A_{0}$, which leads to
\begin{equation}
\Omega_{D}^{2}=\Omega_{0}^{2}+\frac{4\pi\sqrt{\mu(\lambda+2\mu)}}{\rho A_{0}}
\label{UPB}%
\end{equation}
where $A_{0}$ is the unit-cell area. $\Omega_{0}=\Omega_{\mathbf{q}=0}$ is the
temperature- and field-dependent sliding phonon gap related to the
polarization reversal (see below). Carrying out the integral in Eq.~(\ref{OSC}%
) leads to
\begin{equation}
\langle\hat{\xi}_{s}^{2}\rangle=\frac{k_{B}T}{\pi\sqrt{\mu(\lambda+2\mu)}}%
\ln\frac{\sinh\frac{\hbar\Omega_{D}}{2k_{B}T}}{\sinh\frac{\hbar\Omega_{0}%
}{2k_{B}T}}\equiv f\left(  \langle\hat{u}_{s}\rangle,T\right)  . \label{OSC2}%
\end{equation}

A real $\Omega_{0}$ demands that a physically stable phase of the system should fulfill the condition $3\langle \hat{u}_{s}\rangle^2+3f-a_{0}^{2}>0$. When $E=0,$ from Eq.~(\ref{uT}) we have the paraelectric $\langle\hat{u}_{s}\rangle=0$ and ferroelectric $\langle\hat{u}_{s}\rangle=\pm (a_{0}^{2}-3f)^{1/2}$ states for $3f>a_{0}^{2}$ and $3f\leq a_{0}^{2}$, respectively; When $E\neq 0$, always $\langle\hat{u}_{s}\rangle\neq 0$ (see Eq.~(\ref{uT})) and 
\begin{equation}
\langle\hat{u}_{s}\rangle^{2}=a_{0}^{2}-3f\left(  \langle\hat{u}_{s}%
\rangle,T\right)  +\frac{ZEa_{0}^{4}}{4\Delta\langle\hat{u}_{s}\rangle}
\label{SC2}
\end{equation}
which coincides with the ferroelectric case when $E=0$ but $\langle \hat{u}_{s}\rangle\neq 0$. The gap of the sliding phonons under the different conditions are
\begin{equation}
\Omega_{0}=\frac{2}{a_{0}^{2}}\sqrt{\frac{2\Delta}{\rho}}\left\{
\begin{array}
[c]{c}%
(3f-a_{0}^{2})^{1/2},\\
\left[ 2\langle\hat{u}_{s}\rangle^{2}+\frac{Za_{0}^{4}E}{4\Delta\langle
\hat{u}_{s}\rangle}\right]^{1/2},
\end{array}
\begin{array}
[c]{c}%
E=0, \langle\hat{u}_{s}\rangle=0\\
\text{otherwise}\\
\end{array}
\right. \label{Omega}
\end{equation}

In the ferroelectric phase without the field (i.e., $E=0$ and $\langle\hat{u}_{s}\rangle\neq 0$), Eq.~(\ref{Omega}) implies that $\Omega_{0}$ softens with increasing temperature by the average amplitude $|\langle\hat{u}_{s}\rangle|$ but
then increases with temperature in the paraelectric phase via $(3f-a_{0}^{2})^{1/2}$, indicating a dip in $\Omega_{0}(T)$ at the Curie temperature ($T_{c}$). We shall show that this softening leads to an abnormal specific heat at $T_{c}$.

In the following, we solve Eq.~(\ref{SC2}) self-consistently together with
Eq.~(\ref{UPB}) and Eq.~(\ref{Omega}). Its first term represents the spontaneous sliding in the absence of fluctuations that according to the second term is reduced by thermal and zero-point fluctuations. The last term in Eq.~(\ref{SC2}) is the
Stark effect.

\emph{Spontaneous ferroelectricity.} We investigate the spontaneous sliding ferroelectrics without an external field. At zero temperature, the ferroelectricity persists only when the zero-point fluctuations do not destroy the order, i.e.,
$\langle\hat{u}_{s}\rangle^{2}=a_{0}^{2}-3f\left(  \langle\hat{u}_{s}%
\rangle,T=0\right)  >0$, which leads to the condition
\begin{equation}
\gamma\equiv\frac{\hbar}{(\rho A_{0})^{1/2}[\mu(\lambda+2\mu)]^{1/4}a_{0}^{2}%
}<\frac{\sqrt{\pi}}{3} \label{SLQ}%
\end{equation}
that does not require ferroelectricity and holds for any sliding structural phase transitions. The parameter $\gamma$ measures the ratio of the mean-square amplitude of zero-point fluctuations to the squared distance between minimum energy states. $\gamma=\sqrt{\pi}/3$\ marks a quantum phase transition and when $\gamma>\sqrt{\pi}/3$ a quantum paraelectric state occurs as in SrTiO3 and KTaO3 \cite{Muller1979,Rowley2014,Spaldin2021}. Eq.~(\ref{SLQ}) states that bilayers with large unit-cell mass ($\rho A_{0}$), high intralayer stiffness and a large distance between sliding minima favour the order. According to Table~\ref{Table-1}, the zero-point fluctuations are not important for WTe$_{2}$ and h-BN bilayers, as expected. At any finite temperatures, $\Omega_{0}\rightarrow 0$ and $f(\langle\hat{u}_{s}%
\rangle,T\neq 0)\rightarrow\infty$ when $\Delta\rightarrow 0$, which implies the absence of order as follows from the Mermin-Wagner theorem \cite{Mermin1966}. Here we predict a stricter condition for a sliding phase transition, \textit{viz}. not only $\Delta>0$ but also $\gamma<\sqrt{\pi}/3$.

We next address the thermal dynamics of robust sliding ferroelectrics such as
WTe$_{2}$ and h-BN bilayers, in which $\gamma\ll\sqrt{\pi}/3$. At low
temperatures $k_{B}T\ll\hbar\Omega_{0}$ and $E=0,$ the small fluctuations $(3f)$ on the right-hand side of Eq.~(\ref{SC2}) may be
approximated by $\langle\hat{u}_{s}\rangle\approx\pm a_{0}[1-3f(a_{0}%
^{2},T)/(2a_{0}^{2})]$, which leads to
\begin{equation}
\langle\hat{u}_{s}(T)\rangle=\langle\hat{u}_{s}(0)\rangle-\frac{3k_{B}T}%
{2\pi\sqrt{\mu(\lambda+2\mu)}a_{0}}\exp\left(  -\frac{\hbar\Omega_{0}}{k_{B}%
T}\right)\label{LOW}
\end{equation}
where $\langle\hat{u}_{s}(0)\rangle\approx \pm a_{0}\left[  1-3\gamma/(2\sqrt{\pi
})\right]$. In a ferroelctric ($Z\neq 0$) the associated pyroelectric coefficient reads
\begin{equation}
\frac{\partial\langle P(T)\rangle}{\partial T}=-\frac{3k_{B}Z}{2\pi\sqrt
{\mu(\lambda+2\mu)}a_{0}}\frac{\hbar\Omega_{0}}{k_{B}T}\exp\left(
-\frac{\hbar\Omega_{0}}{k_{B}T}\right)  \label{LOWPY}, 
\end{equation}
which differs from the $T^{-1/2}$ prefactor found for 3D ferroelectrics
\cite{Glass1976,Tang2022a}. Eq.~(\ref{LOW}) predicts reduced polarization at thermal energies far below the sliding phonon gap $\Omega_{0}\approx (4/a_{0})(\Delta
/\rho)^{1/2}$.

\begin{figure}[ptb]
\centering
\par
\includegraphics[width=8.6 cm]{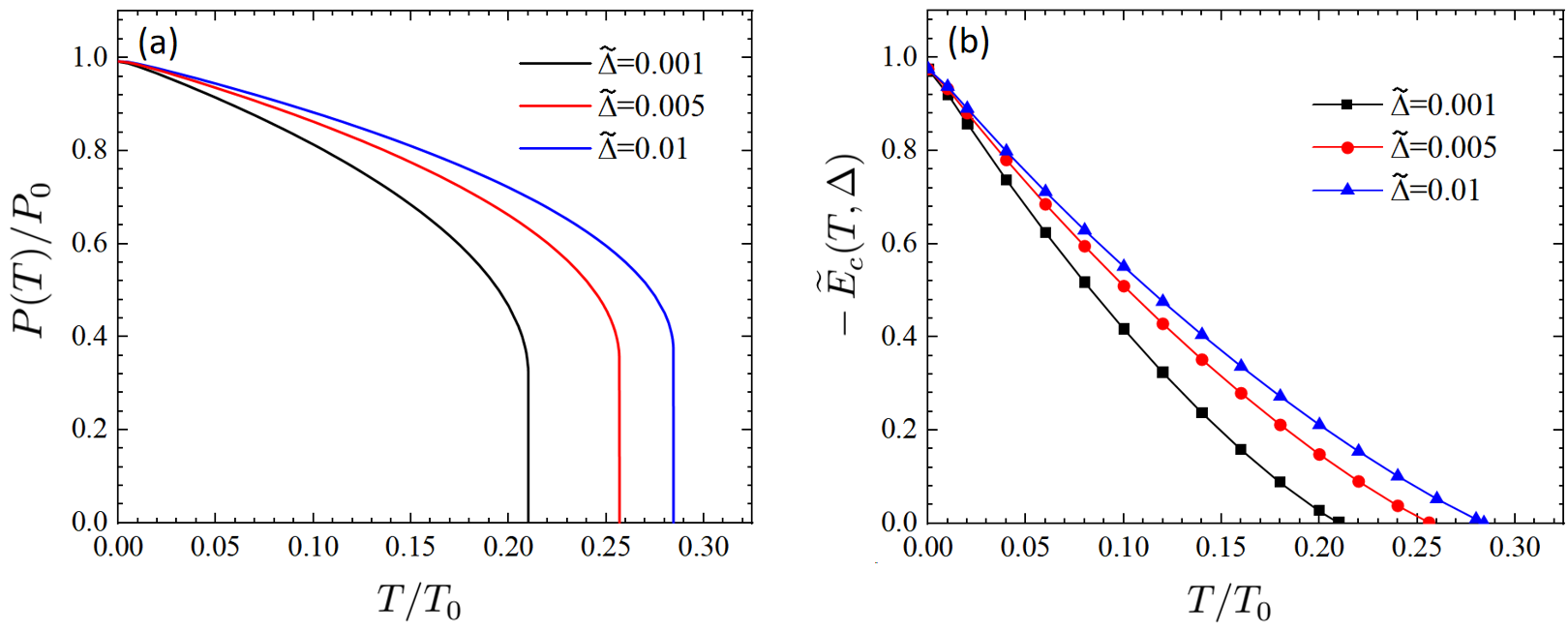}\newline\caption{(a) The
polarization $P(T)$ and (b) the critical switching field $-\widetilde{E}_{c}$ normalized by $8\Delta/(3\sqrt{3}P_{0})$ as a function of temperature for various dimensionless switching barriers $\widetilde{\Delta}=A_{0}\Delta/k_{B}T_{0}$, where we adopt $\gamma=0.01$.}%
\label{Fig-2}%
\end{figure}

Higher temperatures and larger fluctuations increasingly reduce the
polarization. $\langle\hat{u}_{s}\rangle$ does not vanish until the infrared
divergence of $\lim_{\langle\hat{u}_{s}\rangle\rightarrow0}f(\langle\hat
{u}_{s}\rangle,T\neq0),$ i.e., the critical fluctuations signals the phase
transition, which indicates a first-order sliding phase transition, see Fig.~\ref{Fig-2}(a). We estimate the Curie temperature $T_{c}$ by the condition
$\lim_{T\rightarrow T_{c}^{-}}\partial\langle\hat{u}_{s}\rangle/\partial
T\rightarrow\infty$. $T_{c}$ solves Eq.~(\ref{SC2}) with $E=0$
\begin{align}
2\langle\hat{u}_{s}\rangle+3\frac{\partial f\left(  \langle\hat{u}_{s}%
\rangle,T_{c}\right)  }{\partial\langle\hat{u}_{s}\rangle}  &  =0\\
\langle\hat{u}_{s}\rangle^{2}-a_{0}^{2}+3f(\langle\hat{u}_{s}\rangle,T_{c})
&  =0.
\end{align}
In h-BN and WTe$_{2}$ bilayers $\hbar\Omega_{D}\ll k_{B}T_{c}\ll\mu(\lambda
+2\mu)a_{0}^{4}/(A_{0}\Delta)$ such that
\begin{align}
&  T_{c}=\frac{2\pi T_{0}}{3(1+\ln[1+\pi^{2} T_{0}^{2}/(6T_{c}T_{\Delta})])} \label{Tc}\\
&  \left.  \langle\hat{u}_{s}\rangle\right.  \vert_{T=T_{c}^{-}}=\pm a_{0} \sqrt{\frac{3T_{c}%
}{2\pi T_{0}}}
\end{align}
where $k_{B}T_{0}=\sqrt{\mu(\lambda+2\mu)}a_{0}^{2}$ is a measure of the
energy cost of flipping an individual local dipole while $k_{B}T_{\Delta}=A_{0}\Delta$ is the
barrier per unit cell when switching the entire polarization coherently. The
predicated first-order phase transition agrees with the conclusion for R-stacked WSe2 bilayer \cite{Liu2022}. 

\emph{Intrinsic switching field.} Shorted metallic gates such as graphene on
both sides of the ferroelectric screen the ferroelectrict dipoles, while a
voltage bias generates the electric field $E$ in Eq.~(\ref{Hal}). The
screening modifies the electrostatic interactions and stabilizes a single domain configuration compared to the ungated
situation, but otherwise does not affect the physics. According to
Eq.~(\ref{SC2}) the (non-linear) ferroelectric susceptibility
\begin{align}
\chi(T,E)  &  =Z\frac{\partial\langle\hat{u}_{s}(T,E)\rangle}{\partial
E}\nonumber\\
&  =\frac{Z^{2}a_{0}^{4}}{8\Delta\langle\hat{u}_{s}\rangle^{2}}\left[
1+\frac{3}{2\langle\hat{u}_{s}\rangle}\frac{\partial f}{\partial\langle\hat
{u}_{s}\rangle}+\frac{Za_{0}^{4}E}{8\langle\hat{u}_{s}\rangle^{3}\Delta
}\right]  ^{-1}. \label{SUS}%
\end{align}

A large external field against the polarization destabilizes the ferroelectric
order by decreasing the phonon gap until it switches at a coercive field
$E_{c}$ determined by $\lim_{E\rightarrow E_{c}}\chi(T,E)\rightarrow\infty:$
\begin{equation}
E_{c}(T,\Delta)=-\frac{8\Delta\langle\hat{u}_{s}\rangle_{c}^{3}}{Za_{0}^{4}%
}\left[  1+\frac{3}{2\langle\hat{u}_{s}\rangle_{c}}\frac{\partial
f(\langle\hat{u}_{s}\rangle_{c},T)}{\partial\langle\hat{u}_{s}\rangle_{c}%
}\right]  . \label{EC1}%
\end{equation}
where $\langle\hat{u}_{s}\rangle_{c}$ follows from Eq.~(\ref{SC2}) for
$E=E_{c}$, i.e.,
\begin{equation}
3\langle\hat{u}_{s}\rangle_{c}^{2}=a_{0}^{2}-3\left[  f(\langle \hat{u}_{s}\rangle_{c}, T)+\langle\hat{u}%
_{s}\rangle_{c}\frac{\partial f(\langle\hat{u}_{s}\rangle_{c},T)}%
{\partial\langle\hat{u}_{s}\rangle_{c}}\right]  . \label{EC2}%
\end{equation}
Since $\partial f(\langle\hat{u}_{s}\rangle,T)/\partial\langle\hat{u}%
_{s}\rangle^2<0$%
\begin{equation}
- E_{c} <\frac{8\Delta\langle\hat{u}_{s}\rangle_{c}^{3}%
}{Za_{0}^{4}}\equiv-E_{0}  \label{E0}%
\end{equation}
where $\Omega_{0}\left(T, E_{0}\right)  =0$. The ferroelectric order therefore
switches before the gap vanishes, in contrast to bulk ferroelectrics in which
$\Omega_{0}\left(  E_{c}\right)  =0,$ i.e., at relatively low coercive fields
in spite of the high thermal stability.

Figure~\ref{Fig-2}(b) displays numerical solutions of Eq.~(\ref{EC1}) and
Eq.~(\ref{EC2}) for $E_{c}(T,\Delta)$ as a function of temperature for various
$\Delta$ with $E_{c}$ normalized by the classical switching field $8\Delta
/(3\sqrt{3}P_{0})$ in the absence of any fluctuations. $E_{c}(T,\Delta)$ is well
fitted by the power law
\begin{equation}
E_{c}(T,\Delta)=-\frac{8\Delta}{3\sqrt{3}P_{0}}\left(  1-\frac{3\gamma}%
{\sqrt{\pi}}\right)  ^{3/2}\left(  1-\frac{T}{T_{c}}\right)  ^{\eta}%
\end{equation}
The first term in brackets on the r.h.s. represents the effect of quantum
fluctuations. The second one is a Curie-Weiss Law with fitted critical
exponent $\eta\approx1.35$, which is slightly smaller than that of bulk
ferroelectrics with a second-order phase transition ($\eta=1.5$)
\cite{Ducharme2000}. $E_{c}$ is real when the ferroelectric order is stable,
i.e. when $\gamma<\sqrt{\pi}/3$ and $T<T_{c},$ as it should. The above
coercive field  holds for the coherent switching
of a single ferroelectric domain \cite{Ducharme2000, Fridkin2001} and is of order $\sim1-10\,\,\mathrm{G}$V/m for WTe$_2$ and h-BN bilayers. This number
is an order of magnitude larger than observed switching fields
\cite{Fei2018a, Yasuda2021, Stern2021}. Structural disorder such as
dislocations and twisting should reduce the switching field, but their
modelling is beyond the scope of the present paper.

\begin{figure}[ptb]
\centering
\par
\includegraphics[width=6.6 cm]{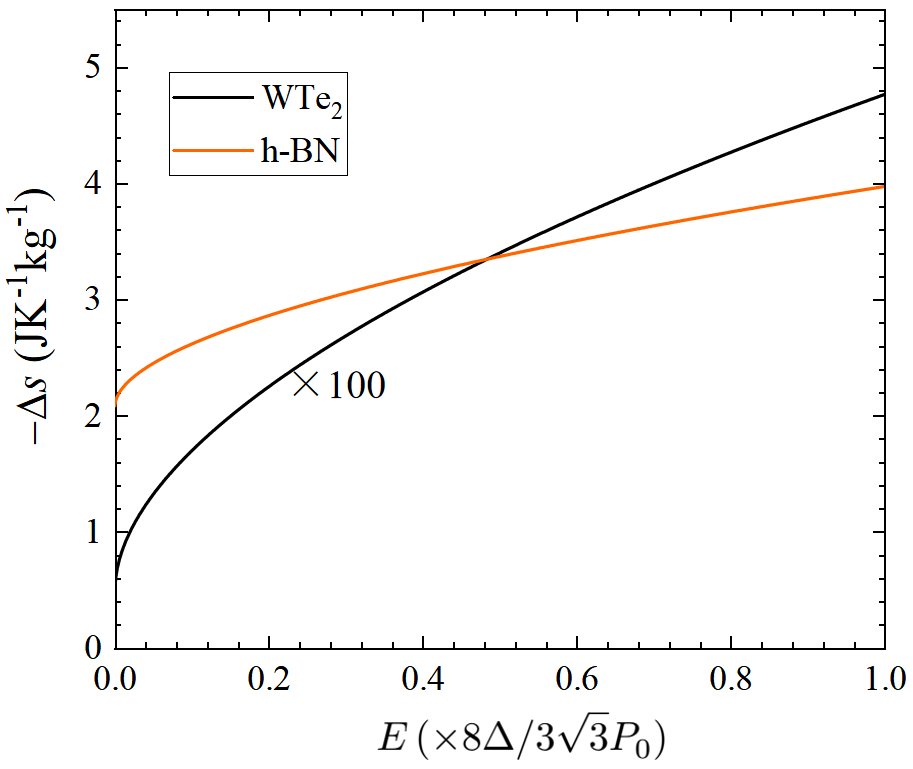}\newline\caption{The entropy change per
unit mass $\Delta s(E)$ with the electric field at $T=T_{c}^{+}$,
where the field is in unit of $8\Delta/(3\sqrt{3}P_{0})$. $\lim_{E\rightarrow 0^+}T_{c}\Delta s(E)$ corresponds to the latent heat generated by the polarization of disordered dipoles. }%
\label{Fig-3}%
\end{figure}

\emph{Electrocaloric effect and specific heat.} The electrocaloric effect
refers to temperature changes caused by the adiabatic (de)polarization of the
ferroelectric order by applied electric fields. The effect is especially large
around first-order phase transitions and interesting for heat management
applications \cite{Liu2016}. The entropy of an ensemble of non-interacting
bosons reads
\begin{align}
S(T,E)=  &  k_{B}\sum_{\mathbf{q}}\left[  (1+n_{\mathbf{q}})\ln
(1+n_{\mathbf{q}})-n_{\mathbf{q}}\ln n_{\mathbf{q}}\right] 
\end{align}
where $n_{\mathbf{q}}=\{\exp[\hbar\Omega_{\mathbf{q}}/(k_{B}T)]-1\}^{-1}$ is the
Planck distribution of the sliding phonons. The isothermal field derivative of entropy then reads
\begin{align}
\frac{\partial S(T, E)}{\partial E}=  &  -\frac{\rho Ak_{B}%
}{4\pi\sqrt{(\lambda+2\mu)\mu}}\frac{\partial\Omega_{0}^{2}}{\partial
E}\int_{x_{0}}^{x_{D}}\frac{xe^{x}}{(e^{x}-1)^{2}%
}dx \label{SEN}%
\end{align}
where $A$ is the area of bilayer and $x_{0(D)}=\hbar\Omega_{0(D)}/(k_{B}T)$. Figure~\ref{Fig-3} shows the entropy change $\Delta s(E)$ (per unit mass) as a function of the
external electric field at $T=T_{c}^{+}$ for WTe$_{2}$ and h-BN bilayers,
where $\lim_{E\rightarrow 0^+}T_{c}\Delta s(E)$ corresponds to the latent heat freed by the polarization of the dipoles. $\Delta s(E)$ is significant
for the h-BN bilayer being of the order of $\mathrm{JK}^{-1}\mathrm{kg}^{-1}$,
but two orders of magnitude smaller in the WTe$_{2}$ bilayer.

For temperature-independent Lam\'{e} parameters, the specific heat of the sliding phonons at a fixed electric field reads
\begin{align}
C_{E}=  &  T\frac{\partial S}{\partial T}=\frac{\rho
Ak_{B}^{3}T^{2}}{2\pi\hbar^{2}\sqrt{(\lambda+2\mu)\mu}}\int_{x_{0}}^{x_{D}%
}\frac{x^{3}e^{x}}{(e^{x}-1)^{2}}dx\nonumber\\
-  &  \frac{\rho Ak_{B}T}{4\pi\sqrt{(\lambda+2\mu)\mu}}\frac
{\partial\Omega_{0}^{2}}{\partial T}\int_{x_{0}}^{x_{D}}%
\frac{xe^{x}}{(e^{x}-1)^{2}}dx. \label{CE}%
\end{align}
The first term in Eq.~(\ref{CE}) follows from the conventional 2D Debye model, while the second one reflects the softening of $\Omega_{0}$ and is singular at the phase transition since $\left. \partial\Omega_{0}^{2}/\partial T\right\vert _{T=T_{c}^{-}}%
\propto\left.  \partial\langle\hat{u}_{s}\rangle^{2}/\partial T\right\vert
_{T=T_{c}^{-}}\rightarrow \infty$. This divergent specific heat might be observed in the associated anomalous heat transport that is beyond
the scope of our paper.

\emph{Discussion.} We can compare the sliding ferroelectricity with 2D
magnetism. In contrast to usual magnets, the zero-point fluctuations explicitly
reduces the sliding ferroelectric order and Curie temperature, because in magnetic systems quantum spins rather than classical magnetic dipoles order and a nonvanishing magnon gap is already a sufficient condition\ for an phase
transition~\cite{Superpara}. Otherwise, at low temperatures Eq.~(\ref{LOW})
resembles the magnetization of 2D ferromagnets as limited\ by magnon
excitations \cite{Bruno1991}. Here the polarization decreases with temperature
due to \textquotedblleft ferrons\textquotedblright, i.e. phonon\ excitations
that carry electric dipoles \cite{Bauer2021, Tang2022a, Tang2022b}. We find an
explicit expression for the reduction of the classical ground state
polarization $Za_{0}$ by zero-point as well as thermal fluctuations. At
sufficiently low temperatures $\langle P(T)\rangle\simeq Z a_{0}[1-3/(2a_{0}%
^{2})f(a_{0}^{2},T)],$ hence the electric dipole carried by a single sliding phonon with wave vector $\mathbf{q}$ is $\delta p_{\mathbf{q}}=-3\hbar Z/(\rho
a_{0}\Omega_{\mathbf{q}})$. We can rewrite Eq.~({\ref{LOW}}) as $\left\langle
P(T)\right\rangle =\left\langle P(0)\right\rangle -\int d^{2}\mathbf{q}%
/(2\pi)^{2}\delta p_{\mathbf{q}}n_{\mathbf{q}}$.

Eq.~(\ref{Tc}) is similar to that of the 2D magnets after replacing the
exchange interaction by $\sqrt{\mu(\lambda+2\mu)}a_{0}^{2}$ or $k_{B}T_{0}$
\cite{Bander1988,Lado2017}. We now understand the stability of sliding
ferroelectricity in terms of the high intralayer stiffness that governs the
energy scale needed to destroy its order $k_{B}T_{0}$ ($\sim0.1-1\,$eV), which is much larger than the 2D magnetic exchange
interaction ($\lesssim10$ meV). The estimates of the critical temperatures in
Table~\ref{Table-1} $T_{c}=660\,$K $\left( T_{c}=1.58\times10^{4}
\,\text{K}\right)$ for WTe$_{2}$ $\left(  \text{h-BN}\right)  $ bilayers agree qualitatively with experiments that report $T_{c}\sim 350\,$K for WTe$_{2}$ \cite{Fei2018a} and a nearly temperature-independent polarization of the BN bilayer in a wide
temperature range up to room temperature \cite{Yasuda2021}.

The present minimal model of sliding phase transitions can be extended and
improved by numerical modelling. Here we consider only unidirectional lateral
sliding, which is analogous to a one-component polarization approximation in
the Landau-Ginzburg-Devonshire theory \cite{Chandra2007}. We may refine this
model by including the coupling with other degrees of freedom, e.g., flexural
and transverse in-plane displacements. The continuum mechanics is not accurate
when the temperatures exceed the Debye temperature and should be checked by lattice
dynamical calculations. Disorder can give rise to position-dependent switching
fields and stick-slip domain formation. The structural stability of twisted
states that generate Moir\.{e} patterns in van der Waals bilayers can be addressed by an appropriate generalization for transitions that involve small twist angles \cite{Peymanirad2017,Bagchi2020,
Zhu2021}. 

\emph{Conclusion:} We model the thermodynamics of 2D sliding ferroelectrics
driven by an external field in a continuum mean-field approximation. We
explain the high Curie temperatures of recently discovered ferroelectrics in spite of ultralow switching fields. We
predict a critical specific heat and a scaling law between the cohesive electric field and temperature. The combination of ultralow switching field and high $T_{c}$ endows the 2D sliding
ferroelectrics with unique functionality for potential applications in
high-integration nanoelectronics.

\emph{Acknowledges:} P. T. and G. B. are
supported by JSPS KAKENHI Grant No. 19H00645 and G. B. is also supported by 22H04965.


\end{document}